\documentstyle[aps,psfig,epsfig,multicol,eqsecnum]{revtex}

\begin{document}
\draft

\title{Wavelets as basis functions in canonical quantization}
\author{M. Havukainen$^1$}
\address{
$^1$Helsinki Institute of Physics, P. O. Box 9, FIN-00014 Helsingin Yliopisto, Finland
}

\date{\today}

\maketitle

\begin{abstract}
Canonical quantization of electromagnetic field is traditionally
done using
plane waves. It is possible to formulate the quantization using other
complete set of basis functions.
Wavelets are a special kind of functions which are localized
in real as well as in Fourier space. In this paper we show how wavelets
can be used as basis functions in canonical quantization. A countable set
of mode functions are obtained.
The general formalism of the change of basis is the same for all
wavelets which satisfy a multiresolution analysis.
\end{abstract}

\begin{multicols}{2}

\section{INTRODUCTION}

Canonical quantization of electromagnetic field is traditionally done
using plane waves. The field is enclosed into a cubic cavity and the
field operators are expanded using eigenfunctions of the cavity, i.e.,
plane waves. After quantization it is possible to take the limit of
infinite cavity and eliminate the unphysical finite size cavity.
The inconvenience of the use of plane waves is that they are
delocalized in real space. This makes it difficult to formulate for
example photodetection theories. Photodetectors measure the field
locally and are sensitive over a finite bandwidth of frequencies.
It is possible to use
any complete set of basis functions in the quantization \cite{mandel,blow},
so a formulation with a more localized basis functions would be desirable.

The theory of multiresolution analysis (MRA) has been under intensive
study during the recent years \cite{burrus,goswami,chui,daubechies}.
A complete set of basis functions in
MRA are called wavelets. Wavelets are localized in real as well as
in Fourier space. They are parameterized by scale and
translation parameters which both get integer values. The translation
parameter translates the wavelet in real space. For large negative
values of the scale parameter the wavelet is wide and for large
scale parameter values narrow. Typical wavelets are orthogonal with
respect to both indices. There are many wavelets with
different characteristics. Differences include how well wavelets
are localized, what kind of Fourier transforms they have, whether they
are real or complex and how symmetric they are. It is
interesting that some wavelets have compact support, i.e., they are
zero outside a certain finite length interval.
These kind of wavelets do not have analytical expressions.
The main theory of multiresolution
analysis is the same for all useful wavelets.

In this paper we show
how wavelets can be used as basis functions in the canonical
quantization. Real and orthonormal wavelets are used.
New mode functions and operators are
linear transforms of plane waves and the corresponding operators.
Different vector valued mode functions for electric and magnetic fields
are obtained. New creation and annihilation operators are the
same for both fields and satisfy bosonic commutation relations.
This means that formalism of the new operators remains the same
and makes it easy to use the new basis. New mode functions are
localized and have similar properties as
wavelets, for examble they are parameterized by scale and translation
parameters.

In Sec. II we give a short introduction to the theory of multiresolution
analysis, scaling functions and wavelets. In Sec. III we derive
equations for field operators in wavelet basis. In Sec. IV
the theory developed in earlier sections is applied to some simple
quantum mechanical simulations and the wavelet
and plane wave bases are compared. Finally in Sec. V we give our
conclusions and suggest several generalizations of the theory developed
in this paper.

\section{Basic properties of scaling functions and wavelets}

\subsection{Multiresolution analysis and wavelets}

In the following we give a brief introduction to the basic properties
of wavelets. The discussion follows books \cite{burrus,goswami,chui}.
We start with the scaling  function
$\phi_l(t)=\phi(t-l), \ l\in Z, \ \phi\in L^2$,
which spans the function space $A_0$

\begin{equation}
f(t)=\sum\limits_l a_l\phi_l(t), \ \ \ \ f(t)\in {A_0}.
\end{equation}
Introducing a new index $s$ by the formula

\begin{equation}
\label{scalingforphi}
\phi_{s,l}(t)=2^{s/2}\phi(2^st-l), \ \ \ \ s,l\in Z
\end{equation}
we can define different function spaces $A_s$.
If $s$ is large, $s\gg 0$,
the scaling function is narrow and peaked around some center point. Large
negative values $s\ll 0$ gives a function which is wide.
The parameter $k$ translates the scaling function in a given scale
by a factor $2^{-s}$.

In order to the scaling function
to satisfy a multiresolution analysis (MRA) the function space
$A_{s+1}$ must include the function space with a lower index $A_s$

\begin{equation}
\label{MRA}
A_s \subset A_{s+1}, \ \ \ \ s \in Z
\end{equation}
When the parameter $s$ approaches infinity we get a space of square
integrable functions $A_{\infty}=L^2$ and when $s$ approaches minus
infinity the result is a null space $A_{-\infty}=\{0\}$.
Because of multiresolution analysis,
it is possible to expand every function in a function space $A_s$
using basis functions of function space $A_{s+1}$. Specifically we
can expand scaling functions $\phi_{s,l}(t)$ in $A_s$ using basis
functions of $A_{s+1}$. This gives

\begin{equation}
\label{MRAforphi}
\phi(t)=\sum\limits_n h(n)\sqrt{2}\phi(2t-n), \ \ \ n \in Z
\end{equation}
with some coefficients $h(n)$. We have chosen $s=0$ and used Eq.
(\ref{scalingforphi}).

The wavelet space is defined to be a difference space between different
function spaces spanned by the scaling functions. We define $W_s$ as

\begin{equation}
\label{division}
A_{s+1}=A_s\oplus W_s.
\end{equation}
Intuitively $W_s$ contains functions which must be added to $A_s$
in order to get $A_{s+1}$. The basis functions in the function space
$W_s$ are called wavelets $\psi(t)$. All functions in $W_s$ can be
expanded using translations of a fundamental wavelet, which are obtained
using the same equation as for scaling functions, Eq. (\ref{scalingforphi}).
Because
$W_s\subset A_{s+1}$, the wavelet function in $W_s$ can be expanded
using scaling functions in $A_{s+1}$. We get

\begin{equation}
\label{MRAforpsi}
\psi(t)=\sum\limits_n h_1(n)\sqrt{2}\phi(2t-n), \ \ \ n \in Z,
\end{equation}
which corresponds to the equation (\ref{MRAforphi}) for $\phi$.
The coefficients $h(n)$ and $h_1(n)$ are not independent. They
satisfy the relation

\begin{equation}
h_1(n)=(-1)^nh(1-n).
\end{equation}

Using Eq. (\ref{MRA}) in Eq. (\ref{division}) we can decompose
the right hand side into subspaces with lower indices. After one
iteration we get

\begin{equation}
A_{s+2}=A_{s+1}\oplus W_{s+1}=A_s\oplus W_s\oplus W_{s+1}.
\end{equation}
We can do the iteration repeatedly and in the infinite limit we get

\begin{equation}
L^2=A_s\oplus W_s\oplus W_{s+1}\oplus ...
\end{equation}
where $A_{\infty}=L^2$ has been used. This means that every function
which belongs to $L^2$ can be expanded as

\begin{eqnarray}
\lefteqn{f(t)=\sum\limits_{l=-\infty}^{\infty}c_{s,l}2^{s/2}\phi(2^st-l)}\\
&\hspace{2.0cm}+&\sum\limits_{s'=s}^{\infty}\sum\limits_{l=-\infty}^{\infty}d_{s',l}2^{s'/2}\psi(2^{s'}t-l).\nonumber
\end{eqnarray}
The parameter $s$ is totally arbitrary. We can take a limit $s\rightarrow -\infty$
and because $A_{-\infty}=\{0\}$ the scaling function space
can be eliminated giving

\begin{eqnarray}
\label{waveletexpansion}
f(t)&=&\sum\limits_{s=-\infty}^{\infty}\sum\limits_{l=-\infty}^{\infty}d_{s,l}\psi_{s,l}(t)\nonumber\\
&=&\sum\limits_{s=-\infty}^{\infty}\sum\limits_{l=-\infty}^{\infty}d_{s,l}2^{s/2}\psi(2^st-l),
\end{eqnarray}
i.e., any $L^2$-function can be expanded using translated and scaled wavelets.
In this paper we use only real orthonormal wavelets which have the
property

\begin{equation}
\label{orthogonalityforwavelets}
\int\limits_{-\infty}^{\infty}\psi_{sl}(t)\psi_{s'l'}(t)dt=\delta_{ss'}\delta_{ll'}.
\end{equation}
Multiplying Eq. (\ref{waveletexpansion}) by parts with $\psi_{s'l'}(t)$ and
integrating $\int\limits_{-\infty}^{\infty}dt$ we get the coefficients

\begin{equation}
d_{s,l}=\int\limits_{-\infty}^{\infty}f(t)\psi_{s,l}(t)dt.
\end{equation}
It follows from the multiresolution analysis that the integral over the
wavelet gives zero, i.e.

\begin{equation}
\int\limits_{-\infty}^{\infty}\psi(t)dt=0.
\end{equation}
This means that wavelets must have some kind of oscillating structure
as the name suggests.

So far we have not defined any wavelets explicitly. There are several methods to
construct wavelets for different purposes. One method is to divide the
Fourier space in such a way that the resulting wavelets fulfill
multiresolution analysis requirements and are orthogonal. The wavelets
obtained using this method can be well localized but do not have
a compact support.
Another method is to derive wavelets based on the filter coefficients.
The filter coefficients of typical wavelets satisfy the fundamental
condition

\begin{equation}
\label{fundamentalcondition}
\sum\limits_nh(2n)=\sum\limits_nh(2n+1)=\frac{1}{\sqrt{2}}
\end{equation}
and are orthogonal

\begin{equation}
\label{orthogonalityofhs}
\sum\limits_nh(n)h(n-2k)=\delta_{k0}.
\end{equation}
If the length of the filter coefficient
sequence is long enough these two conditions do not determine the coefficients
uniquely. The remaining degrees of freedom can be used to give additional
desirable properties to wavelets. Typical choices are to
demand the wavelet or scaling function to be as smooth as possible.
These kind of wavelets can have a compact support, i.e., they are zero
outside a specific region.

\subsection{A few examples of wavelets}

\begin{figure}
\centerline{\psfig{file=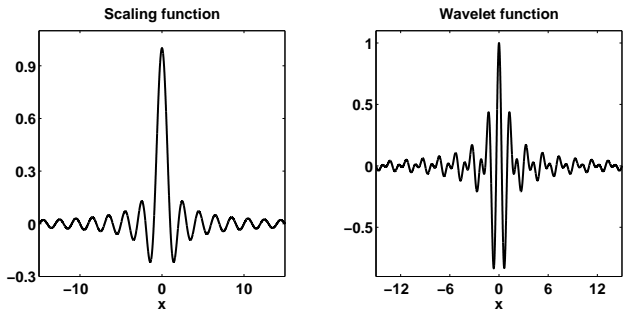,width=7.5cm,bbllx=1cm,bblly=1cm,bburx=21cm,bbury=27cm,angle=90,clip=}}
\vspace{-1cm}
\caption{\narrowtext Fundamental Shannon scaling and wavelet functions ($\phi$ and $\psi$).
Both functions
have analytical expressions, are symmetric and have an oscillating structure.
Oscillations are large also far away from the origin.}
\label{shannonphiandpsi}
\end{figure}

One of the best known scaling function is a sinc-function

\begin{equation}
\phi(x)={\rm sinc}(x)=\frac{\sin(\pi x)}{\pi x}
\end{equation}
with ${\rm sinc}(0)=1$. The corresponding wavelet is $\psi(x)=2\phi(2x)-\phi(x)$.
Figure \ref{shannonphiandpsi} shows the scaling function and the wavelet which are both
localized
around $x=0$. Both functions are not so well localized as
would be desirable for many practical applications. The Fourier
transforms of the scaling and wavelet functions are

\begin{eqnarray}
& & \tilde{\phi}(\omega)=\left\{\begin{array}{ll}
                   \frac{1}{\sqrt{2\pi}}, \ \ \  & \mbox{$\omega<|\pi|$} \\
                   0, \ \ \ & \mbox{otherwise}
               \end{array}
         \right.\\
& & \tilde{\psi}(\omega)=\left\{\begin{array}{ll}
                   \frac{1}{\sqrt{2\pi}}, \ \ \  & \mbox{$\pi<|\omega|<2\pi$} \\
                   0, \ \ \ & \mbox{otherwise.}
               \end{array}
         \right.
\end{eqnarray}
The reason why oscillations do not decay rapidly is
because of
the abrupt changes of the Fourier transforms at $|\omega|=\pi$
and $|\omega|=2\pi$.
One specialty of the sinc-family of wavelets is that the
division of frequency space to different scales is orthogonal, i.e.,
there is no overlap of the Fourier-transforms of wavelets with different
scale parameters $s$. Because of poor localization in real space,
Shannon wavelets are rarely used.
\vspace{-1.5cm}

\begin{figure}
\centerline{\psfig{file=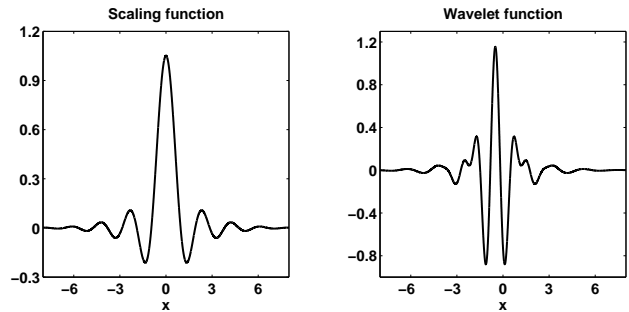,width=7.5cm,bbllx=1cm,bblly=1cm,bburx=21cm,bbury=27cm,angle=90,clip=}}
\vspace{-1cm}
\caption{Fundamental scaling and wavelet functions of Meyer wavelets. Both
functions are better localized than the corresponding Shannon functions
(Fig. \ref{shannonphiandpsi}). Unlike the Shannon functions,
the Meyer scaling function and wavelet are not totally symmetric.}
\label{meyerphiandpsi}
\end{figure}

\begin{figure}
\centerline{\psfig{file=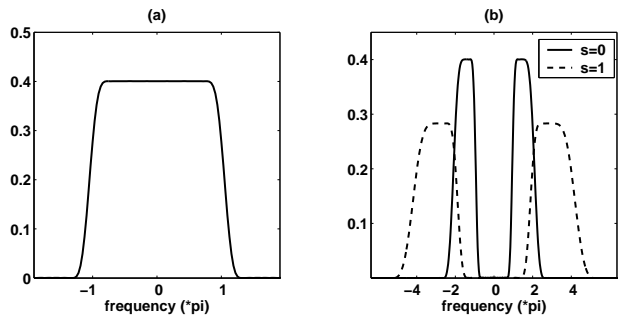,width=7.5cm,bbllx=1cm,bblly=1cm,bburx=21cm,bbury=27cm,angle=90,clip=}}
\vspace{-1cm}
\caption{Left: The absolute value of the Fourier transform of a Meyer scaling
function. The transform is nonzero at small frequencies and decays to zero
at around $|k|=\pi$. The change is smoothened in order to improve
localization. Right: The absolute value of the Fourier transform of a Meyer
wavelet at two different scale parameter values $s=0$ and $s=1$.
There are two peaks symmetrically around zero.
As the scale parameter increases the peaks become lower and are shifted to
higher frequencies.}
\label{meyer1DphiandpsiF}
\end{figure}

It is possible to smoothen the change in Fourier transforms in such a way
that orthogonality and multiresolution analysis requirements are
preserved. The resulting wavelets are called Meyer-wavelets. 
There are several different Meyer scaling and wavelet functions
depending on the smoothening function. The Meyer scaling
and wavelet functions used in this paper are shown in
Fig. \ref{meyerphiandpsi}. They are clearly more localized than
the corresponding Shannon functions.
Absolute values of the Fourier transforms of scaling and wavelet functions
are shown in Fig. \ref{meyer1DphiandpsiF}. The Fourier transform of a scaling
function $\phi$ is flat around $\omega=0$. Around frequencies $\omega=\pm\pi$
the Fourier transform decays to zero and is zero at larger values. The
smoothing compared to Shannon case is clearly seen in both functions.
For wavelets two different scales are shown. As parameter $s$ increases
the transform is wider and shifted further away from the origin.

\vspace{-1.5cm}
\vspace{-1.5cm}

\begin{figure}
\centerline{\psfig{file=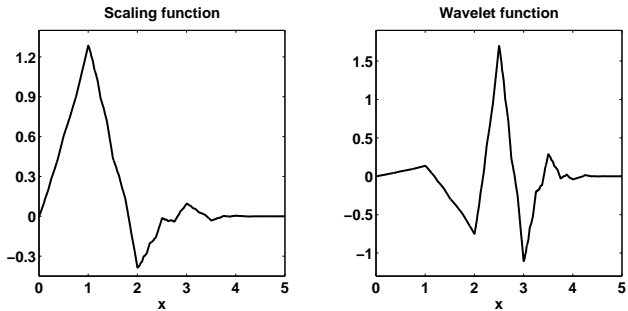,width=7.5cm,bbllx=1cm,bblly=1cm,bburx=21cm,bbury=27cm,angle=90,clip=}}
\vspace{-1cm}
\caption{Daubechies scaling function and wavelet with a filter length
$N=6$. Both functions are continuous and derivable. They have a compact
support, i.e., they are strictly zero outside an interval $0< x< 5$.}
\label{daub6phiandpsi}
\end{figure}

The length of filter sequences $h(n)$ and $h_1(n)$
in Eq. (\ref{MRAforphi}) and (\ref{MRAforpsi}) for sinc and
Meyer scaling fuctions and wavelets are
infinite. As a result of this wavelets do not have a compact
support but are nonzero, although possibly very small, over the
whole $x$-axis.
It is possible to construct wavelets which have compact support, i.e.,
they are exactly zero, not just exponentially small, outside a
specific region. The length of the filter sequence for these
kind of wavelets is finite and the corresponding wavelets and
scaling functions do not have analytical forms.
One group of these wavelets are the Daubechies wavelets. In this
group of wavelets there are several wavelet families which can be
characterized by the lengths of filter sequences. As the length of
the sequence increases the wavelets and scaling functions become
smoother. Figure \ref{daub6phiandpsi} shows Daubechies scaling
and wavelet functions with a filter coefficient length $N=6$
in Eq. (\ref{scalingforphi}) and (\ref{MRAforpsi}). Both
functions are nonzero in the interval $0\leq x\leq 5$. They are
continuous and derivable. The functions are not
symmetric. The general behavior is that the scaling function is
concentrated to the left of the nonzero interval and the wavelet
is significantly nonzero in the middle.
\vspace{-1.5cm}

\subsection{Wavelets in higher dimensions}

The discussion above has been about one dimensional wavelets. We want
to use wavelets as basis functions in canonical quantization so three
dimensional wavelets must be used. One method to construct
multidimensional wavelets is to use
products of one dimensional scaling and wavelet functions \cite{goswami,madych}.
The following products are all wavelets in two dimensions

\begin{eqnarray}
\label{waveletsin2D}
\psi^1({\bf r}) & = & \phi(x)\psi(y)\nonumber \\
\psi^2({\bf r}) & = & \psi(x)\phi(y)\\
\psi^3({\bf r}) & = & \psi(x)\psi(y)\nonumber .
\end{eqnarray}
The integral over all functions above vanishes as is required for wavelets.
The two dimensional scaling function is a product of two one dimensional
scaling functions

\begin{equation}
\phi({\bf r})=\phi(x)\phi(y).
\end{equation}
The wavelets with different scaling and translation parameters become

\begin{equation}
\label{scalingforpsi}
\psi_{s{\bf l}}^i({\bf r})=\psi_{s0}^i({\bf r}-2^{-s}{\bf l})=2^s\psi_{00}^i(2^s{\bf r}-{\bf l}).
\end{equation}
The translation parameter ${\bf l}$ is now a vector with two components which
give the translation
in $x$ and $y$ directions. The scaling parameter is a scalar, so scaling is
the same in both directions.
Note that the scaling of the wavelet is different compared to the one
dimensional case. Any two-dimensional $L^2$-function can be expanded as

\begin{equation}
\label{wexpansionin2D}
f({\bf r})=\sum\limits_{s{\bf l}i}d_{s{\bf l}}^i\psi_{s{\bf l}}^i({\bf r})\equiv\sum\limits_{s=-\infty}^{\infty}\sum\limits_{l_x=-\infty}^{\infty}\sum\limits_{l_y=-\infty}^{\infty}\sum\limits_{i=1}^{3}d_{s{\bf l}}^i\psi_{s{\bf l}}^i({\bf r}).
\end{equation}

\begin{figure}
\centerline{\psfig{file=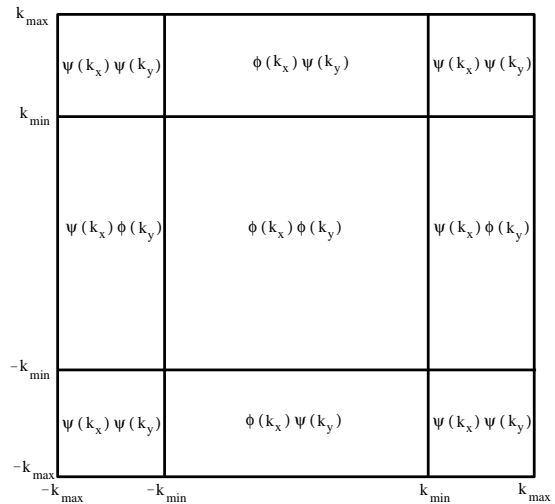,width=7.5cm,bbllx=1cm,bblly=1cm,bburx=21cm,bbury=27cm,angle=0,clip=}}
\vspace{-1.5cm}
\caption{Regions of $k$-space where a two dimensional scaling function
and different wavelets are nonzero at a specific scale. The scaling function
is nonzero at small frequencies. Fourier transforms of
$\psi^1=\phi(k_x)\psi(k_y)$ and $\psi^2=\psi(k_x)\phi(k_y)$
are nonzero at two different regions around $k_y$ and $k_x$ axes.
The Fourier transform of the third wavelet $\psi^3=\psi(k_x)\psi(k_y)$
is nonzero at the corners of the frequency interval.}
\label{regionsin2D}
\end{figure}

Shannon and Meyer wavelets were constructed by dividing the Fourier space.
For these wavelets the three different wavelets in two dimensions,
obtained by multiplying one dimensional scaling functions and wavelets,
are significantly nonzero in different regions of ${\bf k}$-space.
Figure \ref{regionsin2D} shows nonzero regions for three different wavelets.
Wavelets $\psi^1({\bf r})$ and $\psi^2({\bf r})$ are nonzero in regions
$|k_x|\leq k_{min}$,\ \  $k_{min}\leq|k_y|\leq k_{max}$ and 
$|k_y|\leq k_{min}$,\ \  $k_{min}\leq|k_x|\leq k_{max}$ respectively.
Wavelet $\psi^3({\bf r})$ is nonzero in regions
$k_{min}\leq |k_x|\leq k_{max}$, \ \ $k_{min}\leq |k_y|\leq k_{max}$.
With a smaller scaling index $s$, the scaling function in the center is split
into similar regions with different wavelets.
With larger parameter $s$, larger
${\bf k}$-values are divided to regions. The division of frequency space
is exact only for sinc-wavelets. For all other wavelets the Fourier transform
has  nonzero values also outside of its main region.
How well the Fourier transform is localized into the regions depends on
the type of a wavelet used. For wavelets with compact support and small
filter sequence length the division described above is not very well.
\vspace{-1cm}

Three dimensional wavelets can be constructed in a similar way as
two dimensional ones. We get one scaling function and seven
wavelet functions which are products of one dimensional scaling
and wavelet functions

\begin{eqnarray}
\label{waveletsin3D}
\phi({\bf r}) & = & \phi(x)\phi(y)\phi(z) \ \ \ \ \ \psi^1({\bf r}) = \phi(x)\phi(y)\psi(z) \nonumber\\
\psi^2({\bf r}) & = & \phi(x)\psi(y)\phi(z) \ \ \ \ \ \psi^3({\bf r}) = \phi(x)\psi(y)\psi(z) \\
\psi^4({\bf r}) & = & \psi(x)\phi(y)\phi(z) \ \ \ \ \ \psi^5({\bf r}) = \psi(x)\phi(y)\psi(z) \nonumber\\
\psi^6({\bf r}) & = & \psi(x)\psi(y)\phi(z) \ \ \ \ \ \psi^7({\bf r}) = \psi(x)\psi(y)\psi(z). \nonumber
\end{eqnarray}
The division of ${\bf k}$-space is a direct generalization of the two dimensional
case. The scaling relation becomes

\begin{equation}
\label{scalingforpsi3D}
\psi_{s{\bf l}}^i({\bf r})=\psi_{s0}^i({\bf r}-2^{-s}{\bf l})=2^{3s/2}\psi_{00}^i(2^s{\bf r}-{\bf l}).
\end{equation}
The expansion of $L^2$-function is given by Eq. (\ref{wexpansionin2D}) with the
difference that the ${\bf l}$-vector is now three dimensional and index $i$ gets
values from one to seven.

\subsection{Expansion of a plane wave using wavelets}

In the following section we need expansions of plane waves using wavelets.
The normalized plane wave in a box of side $L$ can be expanded as

\begin{equation}
\label{planeexpansion}
\frac{1}{L^{3/2}}e^{i{\bf k}\cdot{\bf r}}=\sum\limits_{s{\bf l}i}d_{{\bf k},s{\bf l}}^i\psi^i_{{s{\bf l}}}({\bf r}).
\end{equation}
Multiplying by parts with $\psi^{i'}_{{s'{\bf l'}}}({\bf r})$ and using orthogonality
of wavelets gives the coefficients

\begin{equation}
\label{d}
d_{{\bf k},s{\bf l}}^i=\frac{1}{L^{3/2}}\int\psi_{s{\bf l}}^i({\bf r})e^{i{\bf k}\cdot{\bf r}}d^3{\bf r}=d_{k_x,sl_x}^{i_x}d_{k_y,sl_y}^{i_y}d_{k_z,sl_z}^{i_z}.
\end{equation}
The factorization is possible because multidimensional wavelets are products
of one dimensional scaling functions and wavelets. The index $i_{x,y,z}$ is either
$0$ or $1$ denoting scaling and wavelet functions respectively.
Similar factorization is possible in Eq. (\ref{planeexpansion}).
Using equation (\ref{scalingforpsi3D}) we get after the change of
integration variable

\begin{equation}
\label{scalingofd}
d_{{\bf k},s{\bf l}}^i=\exp(i2^{-s}{\bf k}\cdot{\bf l})d_{{\bf k},s0}^i=2^{-3s/2}\exp(i2^{-s}{\bf k}\cdot{\bf l})d_{2^{-s}{\bf k},00}^i.
\end{equation}
Thus every coefficient $d^i_{{\bf k},s{\bf l}}$ can be calculated using
the translated and scaled Fourier transform
of a fundamental wavelet. The scaling parameter $s$ changes also the
absolute value of the coefficients whereas the translation parameter ${\bf l}$
gives only a phase shift. Because of orthogonality of plane waves and
wavelets we get the following relations

\begin{eqnarray}
\label{1stsummationoverd}
& & \sum\limits_{s{\bf l}i}d^{i*}_{{\bf k},s{\bf l}}d^i_{{\bf k}',s{\bf l}}=\delta_{{\bf k}{\bf k}'} \\
\label{2ndsummationoverd}
& & \sum\limits_{\bf k}d^{i*}_{{\bf k},s{\bf l}}d^{i'}_{{\bf k},s'{\bf l}'}=\delta_{ss'}\delta_{{\bf l}{\bf l}'}\delta_{ii'}.
\end{eqnarray}

\begin{figure}
\centerline{\psfig{file=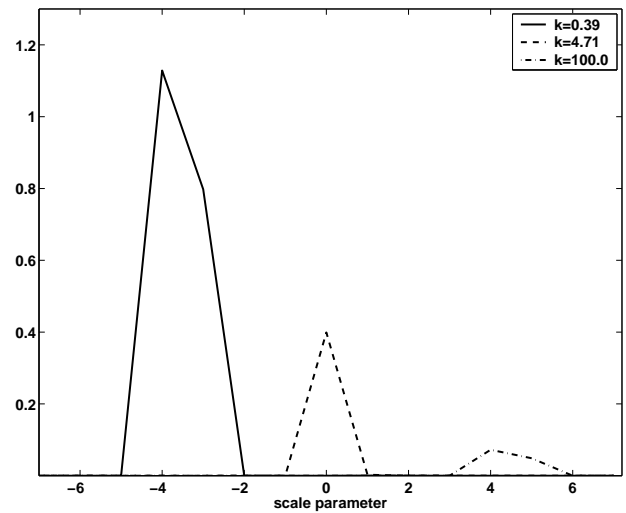,width=7.5cm,bbllx=1cm,bblly=1cm,bburx=21cm,bbury=27cm,angle=90,clip=}}
\vspace{0.2cm}
\caption{The absolute value of the Fourier transform of a wavelet
$d_{k,sl}$ with three different $k$-values. For $k=0.39$
scales $s=-4$ and $s=-3$ are nonzero. Because the value $k=4.71=1.5\pi$ is
in the middle of the scale interval, only one $s$-value is nonzero.
For $k=100.0$ the two scales are nonzero.}
\label{meyerdasfunctionofscale}
\end{figure}

Next we study the parameters $d^i_{{\bf k},s{\bf l}}$ more closely.
We restrict the discussion to one dimension, i.e., the parameters ${\bf k}$
and ${\bf l}$ are one dimensional. Figure \ref{meyerdasfunctionofscale}
shows the absolute values of the coefficients with several different
$k$-values for Meyer wavelets. Note that according to Eq. (\ref{d})
the translation parameter
gives only a phase shift. The coefficients of a plane wave are
shown with three different $k$-values. In all cases coefficients with
only a few scale parameters are nonzero. For small $k$-values $k=0.39$,
two scales
with parameters $s<0$ have nonzero values. Remember that the Fourier
transform of the fundamental ($s=0$) Meyer wavelet is mainly nonzero in the
interval $\pi<k< 2\pi$. The $k$-value $k=4.71=1.5\pi$
is carefully chosen to be in the middle of the scale
$s=0$ interval. It is seen that coefficients for all other $s$-parameters
are zero. With large $k$-values the peaks are centered at a higher scale
as can be seen from the last peak. Peaks with large
scale parameter are smaller than peaks at small scale. This is because
of the factor $2^{-3s/2}$ in equation (\ref{scalingofd}). One could think that
at large scale the peaks are lower because there are more $l$-values
which share the contribution from the single plane wave. Remember that
at high scale values the
shift of a wavelet is smaller when the translation parameter is changed by one unit.
\vspace{0cm}

\section{Canonical quantization using wavelets}

Traditionally canonical quantization of field is done using plane waves
as basis functions \cite{mandel}. The field is enclosed to a cubic of
side $L$. All functions can be expanded using the eigenfunctions of the
quantization volume, i.e., plane waves. The quantization itself is done
by introducing bosonic operators for every basis function. The vector
potential of field can be expanded

\begin{equation}
\hat{\bf A}=\frac{1}{L^{3/2}}\sum\limits_{{\bf k}\sigma}\left(\frac{\hbar}{2\omega_{{\bf k}}\epsilon_0}\right)^{1/2}\left(\hat{a}_{{\bf k}\sigma}{\bf\epsilon}_{{\bf k}\sigma}e^{i{\bf k}\cdot{\bf r}}+{\rm h.c.}\right).
\end{equation}
The vector
$\epsilon_{{\bf k}\sigma}$ gives the polarization of the plane wave with
wave vector ${\bf k}$ and polarization $\sigma$. Operators
$\hat{a}_{{\bf k}\sigma}$ and $\hat{a}_{{\bf k}\sigma}^{\dagger}$
are bosonic annihilation and creation
operators. The parameter ${\bf k}$ gets values

\begin{equation}
{\bf k}=\left(\frac{2\pi n_1}{L},\frac{2\pi n_2}{L},\frac{2\pi n_3}{L}\right),\ \ \ n_{1,2,3}=0,\pm 1,\pm 2...
\end{equation}
For electric and magnetic fields we get

\begin{eqnarray}
\label{operatorE}
\hat{\bf E}({\bf r}) & = & -\frac{\partial\hat{\bf A}({\bf r})}{\partial t}\\
&=&\frac{1}{L^{3/2}}\sum\limits_{{\bf k}\sigma}\left(\frac{\hbar\omega_{\bf k}}{2\epsilon_0}\right)^{1/2}\left(i\hat{a}_{{\bf k}\sigma}\epsilon_{{\bf k}\sigma}e^{i{\bf k}\cdot{\bf r}} + {\rm h.c.}\right) \nonumber\\
\label{operatorB}
\hat{\bf B}({\bf r}) & = & \nabla\times\hat{\bf A}({\bf r})\\
&=&\frac{1}{L^{3/2}}\sum\limits_{{\bf k}\sigma}\left(\frac{\hbar}{2\epsilon_0\omega_{\bf k}}\right)^{1/2}\left(i\hat{a}_{{\bf k}\sigma}({\bf k}\times\epsilon_{{\bf k}\sigma})e^{i{\bf k}\cdot{\bf r}}+ {\rm h.c.}\right).\nonumber
\end{eqnarray}
Field operators above are in plane wave basis, which is parameterized by
${\bf k}$-vectors.

Next we change the basis from plane waves to wavelets.
We could just expand the plane waves using wavelets and use this
expansion in Eqs. (\ref{operatorE}) and (\ref{operatorB}).
However this approach would not allow the separation of mode functions
and operators. Therefore
we proceed by inserting a unity in the form
$\sum\limits_{\bf k}\delta_{{\bf k}{\bf k}'}$ 
to expansions (\ref{operatorE}) and (\ref{operatorB}). The sum is now
divided in such a way that all other ${\bf k}$-values except the
operator ${\bf k}$ are changed to ${\bf k}'$. After the use of Eq.
(\ref{1stsummationoverd}) for the delta function we get

\end{multicols}

\widetext

\begin{eqnarray}
\hat{\bf E}({\bf r}) & = & \frac{1}{L^{3/2}}\sum\limits_{{\bf k}\sigma}\left(\frac{\hbar\omega_{\bf k}}{2\epsilon_0}\right)^{1/2}\left(i\hat{a}_{{\bf k}\sigma}\epsilon_{{\bf k}\sigma}e^{i{\bf k}\cdot{\bf r}} + {\rm h.c.} \right) \\
& = & \frac{1}{L^{3/2}}\sum\limits_{s{\bf l}i}\sum\limits_{{\bf k}{\bf k}'\sigma}\left(\frac{\hbar\omega_{\bf k'}}{2\epsilon_0}\right)^{1/2}\left(i\hat{a}_{{\bf k}\sigma}\epsilon_{{\bf k'}\sigma}d^{i*}_{{\bf k}',s{\bf l}}d^i_{{\bf k},s{\bf l}}e^{i{\bf k'}\cdot{\bf r}}+ {\rm h.c.} \right) \\
& = & \sum\limits_{s{\bf l}i}\sum\limits_{\sigma}\left(\frac{i}{L^{3/2}}\sum\limits_{{\bf k}'}\left(\frac{\hbar\omega_{\bf k'}}{2\epsilon_0}\right)^{1/2}\epsilon_{{\bf k'}\sigma}d^{i*}_{{\bf k'},s{\bf l}}e^{i{\bf k'}\cdot{\bf r}}\right)\left(\sum\limits_{{\bf k}}d^{i}_{{\bf k},s{\bf l}}\hat{a}_{{\bf k}\sigma}\right) + {\rm h.c.} \\
\label{ueexpansion}
& = & \sum\limits_{s{\bf l}i}\sum\limits_{\sigma}\left(\hat{b}^i_{s{\bf l},\sigma}{\bf u}_{s{\bf l},\sigma}^{iE}({\bf r}) + {\rm h.c.} \right),
\end{eqnarray}
\begin{multicols}{2}
where the new basis functions and operators are

\begin{eqnarray}
\label{ueinwbase}
{\bf u}_{s{\bf l},\sigma}^{iE}({\bf r}) & = &\frac{i}{L^{3/2}}\sum\limits_{\bf k}\left(\frac{\hbar\omega_{\bf k}}{2\epsilon_0}\right)^{1/2}d^{i*}_{{\bf k},s{\bf l}}{\bf\epsilon}_{{\bf k}\sigma}e^{i{\bf k}\cdot{\bf r}} \\
\label{ueinwbase2}
 & = & {\bf u}_{s0}^{iE}({\bf r}-2^{-s}{\bf l})=2^{2s}{\bf u}_{{00},\sigma}^{iE}(2^s{\bf r}-{\bf l}) \\
\label{operatorinwbase}
 \hat{b}^i_{s{\bf l},\sigma} & = &\sum\limits_{{\bf k}}d_{{\bf k},s{\bf l}}^{i}\hat{a}_{{\bf k}\sigma}.
\end{eqnarray}
Similarly for the magnetic field we get

\begin{equation}
\label{ubexpansion}
\hat{\bf B}({\bf r})=\sum\limits_{s{\bf l}i}\sum\limits_{\sigma}\left(\hat{b}^i_{s{\bf l},\sigma}{\bf u}_{s{\bf l},\sigma}^{iB}({\bf r}) + {\rm h.c.} \right),
\end{equation}
where the operator $\hat{b}_{s{\bf l},\sigma}^i$ is given by equation (\ref{operatorinwbase}) and

\begin{eqnarray}
\label{ubinwbase}
{\bf u}^{iB}_{s{\bf l},\sigma}({\bf r}) & = & \frac{i}{L^{3/2}}\sum\limits_{\bf k}\left(\frac{\hbar}{2\epsilon_0\omega_{\bf k}}\right)^{1/2}d^{i*}_{{\bf k},s{\bf l}}({\bf k}\times{\bf\epsilon}_{{\bf k}\sigma})e^{i{\bf k}\cdot{\bf r}}\\
\label{ubinwbase2}
 & = & {\bf u}_{s0}^{iB}({\bf r}-2^{-s}{\bf l})=2^{2s}{\bf u}_{{00},\sigma}^{iB}(2^s{\bf r}-{\bf l}).
\end{eqnarray}
The new mode functions behave in the same way as wavelets
when the indices are changed. The translation parameter ${\bf l}$ translates the
mode functions in three dimensions and the parameter $s$ compresses and stretches
them. Even though the quantization in the wavelet basis is in free space the
operators and mode functions are expanded using a countable set of basis
functions.
On the contrary to plane waves the polarization of the new mode
functions is not constant.
As is the case with wavelets, the integral
of the mode functions over quantization volume vanishes

\begin{equation}
\int {\bf u}_{s{\bf l},\sigma}^{iE}({\bf r})d^3{\bf r}=\int {\bf u}_{s{\bf l},\sigma}^{iB}({\bf r})d^3{\bf r}=0.
\end{equation}
This means that also the mode functions have similar 'waviness' as wavelets.

It is easy to show that the new operators $\hat{b}^i_{s{\bf l},\sigma}$ and
$\hat{b}_{s{\bf l},\sigma}^{i\dagger}$ obey bosonic commutation relations

\begin{eqnarray}
& & \left[b^i_{s{\bf l},\sigma},b^{i'}_{s'{\bf l}',\sigma'}\right]=\left[b^{i\dagger}_{s{\bf l},\sigma},b_{s'{\bf l}',\sigma'}^{i'\dagger}\right]=0, \\
& & \left[b^i_{s{\bf l},\sigma}, b^{i'\dagger}_{s'{\bf l}',\sigma'}\right]=\delta_{ii'}\delta_{ss'}\delta_{{\bf l}{\bf l}'}\delta_{\sigma\sigma'}.
\end{eqnarray}
Because of this the operators act like annihilation
and creation operators for wavelet modes. This means that the formalism
in wavelet basis is the same as with plane wave operators.
Multiplying both sides of Eq. (\ref{operatorinwbase}) by $d^i_{{\bf k},s{\bf l}}$
and performing the sum $\sum\limits_{s{\bf l}i}$ we get with the help of the
orthogonality integral

\begin{equation}
\label{asumb}
\hat{a}_{{\bf k}\sigma}=\sum\limits_{s{\bf l}i}d^{i*}_{{\bf k},s{\bf l}}\hat{b}^i_{s{\bf l},\sigma}.
\end{equation}
Similarly it is possible to expand plane wave basis functions using wavelet mode
functions.

The field Hamiltonian in wavelet basis is obtained as an integral of the
energy density over the quantization volume

\begin{equation}
\label{H_F}
\hat{H}_F=\int\left(\frac{1}{2}\epsilon_0\hat{E}^2({\bf r})+\frac{1}{2\mu_0}\hat{B}^2({\bf r})\right)d^3{\bf r},
\end{equation}
where expansions (\ref{ueinwbase}) and (\ref{ubinwbase}) are used for electric
and magnetic field operators. In the calculation of (\ref{H_F}) the following
integrals are needed (last one is listed only for completeness)

\begin{eqnarray}
\label{integral1}
& & \int {\bf u}_{s{\bf l},\sigma}^{iE*}({\bf r})\cdot{\bf u}_{s'{\bf l}',\sigma'}^{i'E}({\bf r})d^3{\bf r}=\delta_{\sigma\sigma'}\frac{\hbar}{2\epsilon_0}w_{s'{\bf l}',s{\bf l}}^{i',i}\\
\label{integral2}
& & \int {\bf u}_{s{\bf l},\sigma}^{iB*}({\bf r})\cdot{\bf u}_{s'{\bf l}',\sigma'}^{i'B}({\bf r})d^3{\bf r}=\delta_{\sigma\sigma'}\frac{\hbar}{2\epsilon_0c^2}w_{s'{\bf l}',s{\bf l}}^{i',i}\\
\label{integral3}
& & \int {\bf u}_{s{\bf l},\sigma}^{iE*}({\bf r})\cdot{\bf u}_{s'{\bf l}',\sigma'}^{i'B}({\bf r})d^3{\bf r}=0,
\end{eqnarray}
where the real coupling constants of the modes are
\end{multicols}
\begin{eqnarray}
\label{wusingds}
w^{ii'}_{s{\bf l},s'{\bf l}'} = w^{i'i}_{s'{\bf l}',s{\bf l}} & = & \sum\limits_{\bf k}\omega_{\bf k}d^{i*}_{{\bf k},s{\bf l}}d^{i'}_{{\bf k},s'{\bf l}'}\\
\label{wusingpsis}
& = & \frac{1}{(2\pi)^{3}}\sum\limits_{\bf k}\int d^3{\bf r}\int d^3{\bf r}'\omega_{\bf k}\Psi^{i}_{s{\bf l}}({\bf r})\Psi^{i'}_{s'{\bf l}'}({\bf r}')e^{i{\bf k}\cdot({\bf r}'-{\bf r})}.
\end{eqnarray}
\begin{multicols}{2}
Using (\ref{scalingofd}) we get

\begin{eqnarray}
\label{omegaasfunctionofF}
w_{s{\bf l},s'{\bf l}'}^{ii'}&=&\sum\limits_{\bf k}\omega_{\bf k}d_{{\bf k},s0}^{i*}d_{{\bf k},s'0}^{i'}\exp(i{\bf k}\cdot(2^{-s'}{\bf l'}-2^{-s}{\bf l}))\nonumber\\
&=&F_{ss'}^{ii'}(2^{-s'}{\bf l'}-2^{-s}{\bf l}),
\end{eqnarray}
where

\begin{equation}
F_{ss'}^{ii'}({\bf x})=\sum\limits_{\bf k}\omega_{\bf k}d_{{\bf k},s0}^{i*}d_{{\bf k},s'0}^{i'}e^{i{\bf k}\cdot{\bf x}}.
\end{equation}
The identity (\ref{omegaasfunctionofF}) shows that the coupling constant
is a function of the difference of the scaled translation parameters only.
Using equations (\ref{integral1}) and (\ref{integral2}) and the identity

\begin{equation}
\int{\bf u}_{s{\bf l},\sigma}^{iE}({\bf r})\cdot{\bf u}_{s'{\bf l}',\sigma'}^{i'E}({\bf r})d^3{\bf r}=-c^2\int{\bf u}_{s{\bf l},\sigma}^{iB}({\bf r})\cdot{\bf u}_{s'{\bf l}',\sigma'}^{i'B}({\bf r})d^3{\bf r}
\end{equation}
we get after a straightforward calculation for the free field Hamiltonian

\begin{equation}
\label{H_Finwbase}
\hat{H}_F=\sum\limits_{\sigma}\sum\limits_{s{\bf l}i}\sum\limits_{s'{\bf l}'i'}\hbar w_{s{\bf l},s'{\bf l}'}^{i,i'}\hat{b}_{s{\bf l}}^{i\dagger}\hat{b}_{s'{\bf l}'}^{i'}+\frac{\hbar}{2}\sum\limits_{\sigma}\sum\limits_{s{\bf l}i}w_{s{\bf l},s{\bf l}}^{i,i}.
\end{equation}
The modes are coupled with the coupling constant
$w_{s{\bf l},s'{\bf l}'}^{i,i'}$. The diagonal elements
$w_{s{\bf l},s{\bf l}}^{i,i}$ give the energy of the field
state which has only one mode with parameters $s$, ${\bf l}$ and
$i$ excited, i.e., energy of one wavelet quantum.
From Eq. (\ref{wusingds}) and
(\ref{wusingpsis}) it is seen that the coupling is nonzero only if the
corresponding wavelets and Fourier transforms have a nonzero overlap.
Because the wavelets are localized both in real and Fourier space this
means that for majority of mode pairs the coupling is zero. The detailed
structure of the coupling constants $w_{s{\bf l},s'{\bf l}'}^{i,i'}$
depends on the wavelet used. The second term in Eq. (\ref{H_Finwbase})
is the result of the self-coupling of the modes and gives the zero
point energy of the field. As is the case for plane waves the energy
per mode is half of the unit excitation field.

\section{Numerical simulations using wavelet modes}

\subsection{Conventions and wavelets used}

In this section we give examples how to apply the theory developed in the
last section. We do the calculations in two rather than in
three dimensions because visualization is easier and all essential features
are included. Thus the $\bf k$-vector is restricted to xy plane
${\bf k}=k_x{\bf e}_1+k_y{\bf e}_2$ and the polarization vector is
${\bf\epsilon}_{{\bf k}1}=-{\bf e}_3$. Here we exclude the other
polarization ${\bf\epsilon}_{{\bf k}2}$. The cross product in the
expansion of the magnetic field (\ref{operatorB}) becomes
${\bf k}\times{\bf\epsilon}_{{\bf k}1}=-k_y{\bf e}_1+k_x{\bf e}_2$.
The choice of the field configuration considered is the same as in our
earlier paper \cite{cavity2d}.
The mode functions for electric and magnetic fields become
\begin{eqnarray}
u_{s{\bf l},1}^{iE}({\bf r}) & = & -\frac{i}{L}\sum\limits_{\bf k}\left(\frac{\hbar\omega_{\bf k}}{2\epsilon_0}\right)^{1/2}d^{i*}_{{\bf k},s{\bf l}}{\bf e}_3e^{i{\bf k}\cdot{\bf r}} \\
u_{s{\bf l},1}^{iB}({\bf r}) & = & \\
& &\hspace{-1cm}\frac{i}{L}\sum\limits_{\bf k}\left(\frac{\hbar}{2\epsilon_0\omega_{\bf k}}\right)^{1/2}d^{i*}_{{\bf k},s{\bf l}}(-k_y{\bf e}_1+k_x{\bf e}_2)e^{i{\bf k}\cdot{\bf r}},\nonumber
\end{eqnarray}
where the summation over ${\bf k}$ is now over the xy plane. In the following
we choose units such that $\hbar=\epsilon_0=c=1$. In all
examples in this section we use the Meyer wavelets shown in
Fig. \ref{meyerphiandpsi}. Meyer wavelets were designed using smoothened
Fourier transform of Shannon wavelets, so the division of the Fourier-plane
shown in Fig. \ref{regionsin2D} is rather valid. The Fourier transforms
of the mode functions are quite similar to the Fourier transforms of the
corresponding wavelets. In all simulations
the parameters are chosen in such a way that it is necessary to consider only
two scales $s=0$ and $s=1$. This means that the frequencies
$|k_x|<\pi$, $|k_y|<\pi$ and $|k_x|>4\pi$, $|k_y|>4\pi$ are excluded.
One has to note that in two dimensions the scaling of the wavelet
mode functions become

\begin{equation}
{\bf u}_{s{\bf l},\sigma}^{iE,B}({\bf r})={\bf u}_{s0,\sigma}^{iE,B}({\bf r}-2^{-s}{\bf l})=2^{3s/2}{\bf u}_{00,\sigma}^{iE,B}(2^s{\bf r}-{\bf l}),
\end{equation}
instead of the scaling in three dimensions given by Eqs. (\ref{ueinwbase2}) and
(\ref{ubinwbase2}).

\begin{figure}
\centerline{\psfig{file=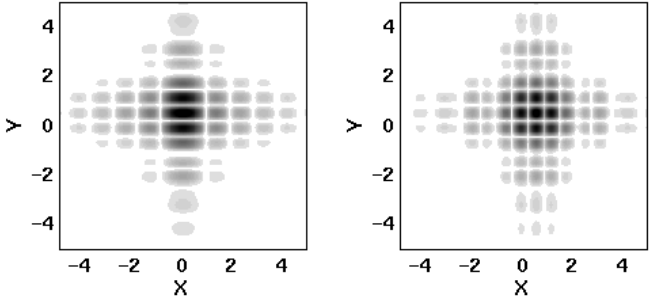,width=4.5cm,bbllx=5cm,bblly=1.8cm,bburx=16cm,bbury=27cm,angle=90,clip=}}
\vspace{0cm}
\caption{Absolute value of electric wavelet mode functions
${\bf u}_{s{\bf l}}^{iE}$ with parameters $s=0$ and ${\bf l}=0$. The
function on the left has index $i=1$ and on the right $i=3$. The
$i=2$ mode function is similar to $i=1$ case with $x$- and $y$-axes changed.
Mode functions are localized at the origin and have wavelet type oscillations.}
\label{meyerue1andue3}
\end{figure}

\begin{figure}
\centerline{\psfig{file=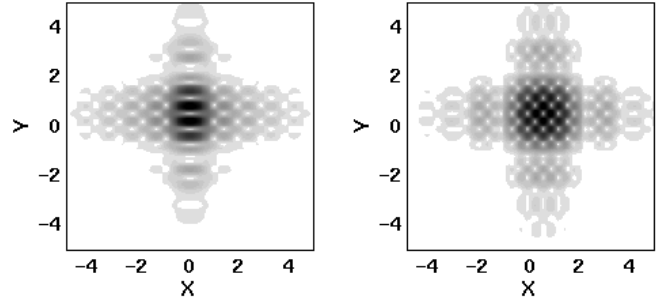,width=4.5cm,bbllx=5cm,bblly=2.3cm,bburx=16cm,bbury=29cm,angle=90,clip=}}
\vspace{0.5cm}
\caption{Absolute value of magnetic wavelet mode functions
${\bf u}_{s{\bf l}}^{iB}$. The parameters are the same as in Fig.
\ref{meyerue1andue3}. Mode functions are localized and different from
the corresponding electric mode functions
(Fig. \ref{meyerue1andue3}).}
\label{meyerub1andub3}
\end{figure}

The absolute values of the electric and magnetic mode functions with indices
$s=0$ and $l_x=l_y=0$ are shown in Figs.
\ref{meyerue1andue3} and \ref{meyerub1andub3}.
On the left in both figures there is the mode function with index $i=1$ and
on the right $i=3$. The $i=2$ mode function is the same as $i=1$ case with
axes $x$ and $y$ changed. All mode functions are clearly localized at the origin
and have wavelet type of oscillating structure. Because the Fourier transform
of $i=3$ two dimensional wavelet is nonzero at higher frequencies than Fourier
transforms of $i=1$ and $i=2$ wavelets, the oscillations of $i=3$ mode function
have smaller details. As was explained earlier
the parameters $s$ and ${\bf l}$ scale and translate mode functions.
\vspace{0cm}
\vspace{0cm}

\subsection{A Gaussian photon}

In this section we study a one photon field which is a superposition of
different single excitation plane wave modes. The distribution of the
absolute values of the mode coefficients is a Gaussian centered
at some point ${\bf k}_0$ with a variance $\Delta_k^2$ in ${\bf k}$-space.
Thus the field state is

\begin{equation}
|\Psi\rangle=\sum\limits_{\bf k}c_{\bf k}|1_{\bf k},\{0\}\rangle=\sum\limits_{\bf k}c_{\bf k}\hat{a}_{\bf k}^{\dagger}|\{0\}\rangle
\end{equation}
where

\begin{equation}
\label{gaussiandistribution}
c_{\bf k}=(2\pi\Delta_{k}^2)^{-1/2}e^{-i{\bf k}\cdot{\bf r}_0}\exp\left(-\frac{({\bf k}-{\bf k}_0)^2}{4\Delta_k^2}\right).
\end{equation}
The transformation to wavelet basis can be done using the expansion of
operators $\hat{a}_{\bf k}$ in a wavelet basis (\ref{operatorinwbase}).
The state vector in wavelet basis becomes

\begin{equation}
|\Psi\rangle=\sum\limits_{s{\bf l}i}c_{s{\bf l}}\hat{b}^{i\dagger}_{s{\bf l}}|\{0\}\rangle,
\end{equation}
where

\begin{equation}
\label{coefficientsinwbase}
c^i_{s{\bf l}}=\sum\limits_{\bf k}d^i_{{\bf k},s{\bf l}}c_{\bf k}.
\end{equation}
Because coefficients $d^i_{{\bf k},s{\bf l}}$ and $c_{\bf k}$ factorize, it is
possible to factorize the sum (\ref{coefficientsinwbase}).

\begin{figure}
\centerline{\psfig{file=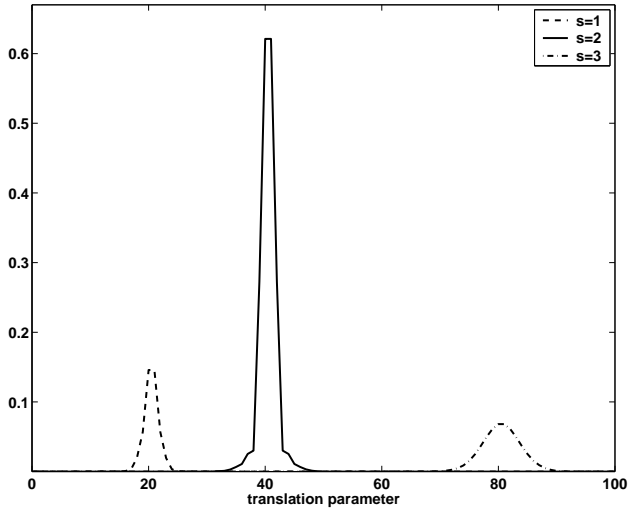,width=7.5cm,bbllx=1cm,bblly=1cm,bburx=21cm,bbury=27cm,angle=90,clip=}}
\vspace{0.2cm}
\caption{Absolute values of wavelet mode coefficients for a one
dimensional Gaussian photon as a function of a translation parameter.
The parameters are $r_0=10.0$, $k_0=18.0$
and $\Delta_k^2=10.0$. Coefficients are zero only for three scales.
The peaks are at different positions because the translation unit
is not the same at different scales. Scale $s=2$ gives the biggest
contribution.}
\label{meyer1Dgaussianstate}
\end{figure}

First we briefly study a one dimensional Gaussian distribution.
We have a Gaussian
photon with parameters $k_0=18.0$, $x_0=10.0$ and $\Delta_k^2=10.0$.
The width of the distribution is relatively large so one would expect
several scales to have nonzero coefficients. Figure
\ref{meyer1Dgaussianstate}
shows coefficients of three different scales as a function of a
translation parameter. The peak on the left has the scale parameter $s=1$.
When the translation parameter is changed by one, the wavelet and mode functions
are translated by $2^{-s}$ in real space. At scale $s=1$ the translation
unit is $2^{-1}=0.5$. The distribution is centered at $x_0=10.0$ so
coefficients at $s=1$ are centered at $l=20$. When $s=2$
and $s=3$ the coefficients around $l=40$ and $l=80$ have nonzero
values. It is seen that a Gaussian state with the parameters used has
the biggest contribution at a scale $s=2$.
\vspace{-0.5cm}

\begin{figure}
\centerline{\psfig{file=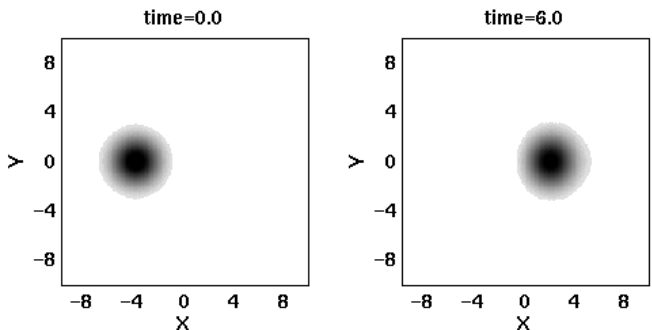,width=4.5cm,bbllx=5cm,bblly=1.8cm,bburx=16cm,bbury=27cm,angle=90,clip=}}
\vspace{0.0cm}
\caption{Time evolution of the energy density distribution of
a Gaussian one photon state with parameters $r_x=-4.0$, $r_y=0.0$,
$k_x=7.0$, $k_y=0.0$ and $\Delta_k^2=0.25$. At $t=0.0$ the
intensity profile is Gaussian. Later at $t=6.0$ the photon has
propagated to the right and the intensity profile is not
perfectly Gaussian.}
\label{gaussianintensity}
\end{figure}

Next we study the time evolution of a two dimensional Gaussian photon.
Figure \ref{gaussianintensity} shows the time evolution of the energy density
distribution of a photon at two different times. The parameters
in Eq. (\ref{gaussiandistribution}) for the initial state at $t=0.0$
are $r_x=-4.0$, $r_y=0.0$, $k_x=7.0$, $k_y=0.0$ and $\Delta_k^2=0.25$.
It is seen that the energy density profile of the single excitation field
which has a Gaussian distribution in Fourier space is also Gaussian.
At a later time $t=6.0$ the photon has propagated to the right and
the intensity profile has spread a little.
The time evolution is qualitatively the same as obtained in
the paper \cite{cavity2d} using plane wave quantization.

\vspace{-0.2cm}

On the contrary to plane wave modes the wavelet modes are coupled
also for the free field. This
makes the operation of the free field Hamiltonian to the statevector,
which is needed in the numerical integration, slower. However, most
of the coupling constants $w_{s{\bf l},s'{\bf l}'}^{i,i'}$ are
zero. Use of this fact makes the operation much faster. The function
$F_{ss'}^{ii'}({\bf x})$ in the calculation of the coupling
coefficients can be used to save memory. In general fewer wavelet
mode functions are needed to represent a localized field state than
plane wave mode functions.

\subsection{Wavelet mode function initial state}

\begin{figure}
\centerline{\psfig{file=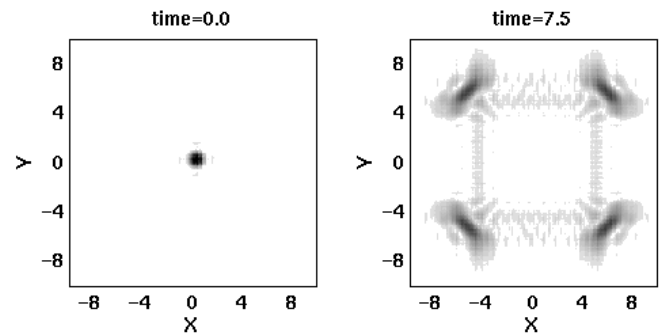,width=4.5cm,bbllx=5cm,bblly=2.5cm,bburx=16cm,bbury=28cm,angle=90,clip=}}
\vspace{0.4cm}
\caption{Time evolution of a state which at time $t=0$ has only one
Meyer wavelet mode excited. The parameters of the mode function are $s=1$,
$l_x=l_y=0.0$ and $i=3$. The intensity profile at $t=0.0$ is localized
at the origin.
At $t=7.5$ the intensity profile has propagated mainly to four
directions away from the center.}
\label{index6intensity}
\end{figure}

Next we study the time evolution of the field state which at time $t=0.0$
has only one wavelet mode excited. The mode which is initially excited
has parameters $s=1$, ${\bf l}=0$ and $i=3$. The intensity of the
field state is shown in Fig. \ref{index6intensity}. It is clearly
localized around ${\bf r}=0$ which is expected based on the parameters
chosen. The Fourier transform for $i=3$ mode function is divided into four
parts at the corners of the frequency interval with a specific scaling
index $s$, as is explained earlier and shown in Fig. \ref{regionsin2D}.
The intensity profile at time $t=7.5$ is shown in
Fig. \ref{index6intensity}. The intensity is divided to four main parts
which all propagate away from the origin. This is understandable based
on the Fourier transform of the mode function.
\vspace{0cm}

\subsection{Decay of a two level atom}

Next we couple a two level atom to the two dimensional field. The interaction
Hamiltonian in dipole approximation can be written as

\begin{equation}
\hat{H}_I=-\hat{\bf D}\cdot\hat{\bf E}({\bf r}_0),
\end{equation}
where $\hat{\bf D}$ is the electric dipole moment operator of the two level atom

\begin{equation}
\hat{\bf D}=(D\hat{\sigma}_++D^*\hat{\sigma}_-){\bf e}_3
\end{equation}
and ${\bf r}_0$ the position of the atom. We take the direction of the
dipole operator to be in the
$z$ direction. Using the wavelet expansion of the field (\ref{ueexpansion}) we
get

\begin{equation}
\hat{H}_I=-\sum\limits_{s{\bf l}i}\sum\limits_{\sigma}(D{\bf u}_{s{\bf l},\sigma}^{iE}({\bf r}_0)\hat{\sigma}_+\hat{b}_{s{\bf l},\sigma}^i+D^*{\bf u}_{s{\bf l},\sigma}^{iE}({\bf r}_0)\hat{\sigma}_-\hat{b}_{s{\bf l},\sigma}^{i\dagger}).
\end{equation}
In the Hamiltonian the rotating wave approximation (RWA) has been used. The
approximation can be done if the mode functions used are well localized in
Fourier space. This is the case with Meyer wavelets. The scale parameter
$s$ determines the frequency interval where the Fourier transform of the
mode function is nonzero. Typically the atom interacts with mode functions
of only a few, maybe one or two, scales.
The mode functions are also spatially localized. Because the interaction
is proportional to the electric mode function evaluated at a position
of the atom, only mode functions which are centered close to the atom
interact with it. In this respect the situation is different compared to
the plane wave mode functions which are delocalized. In general the atom
is coupled to fewer mode functions than in the plane wave quantization.

\begin{figure}
\centerline{\psfig{file=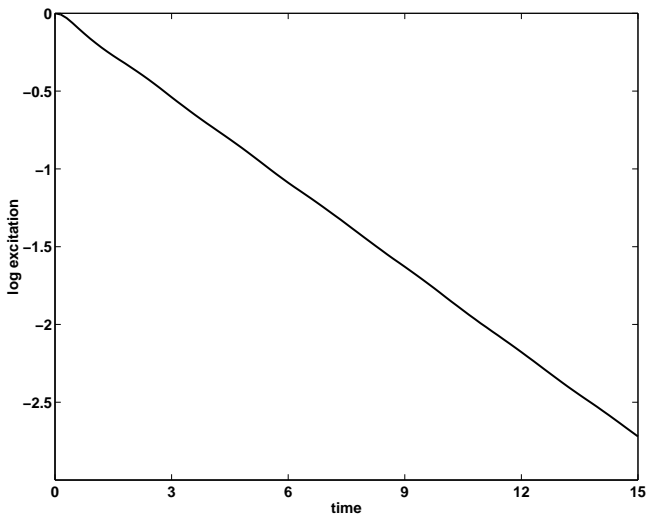,width=7.5cm,bbllx=1cm,bblly=1cm,bburx=21cm,bbury=27cm,angle=90,clip=}}
\vspace{0cm}
\caption{Logarithm of the excitation probability of a decaying two
level atom as a function of time. The resonance frequency of the atom
is $\omega=10.0$ and the dipole constant $D=0.06$. The decay is exponential
with a decay constant $\Gamma=0.18$.}
\label{logatomamplitudes}
\end{figure}

Fig. \ref{logatomamplitudes} shows the logarithm of the excitation probability
of the atom as function of time. The resonance frequency of the atom is
$\omega=10.0$ and the dipole coupling D=0.06. The atom decays energy to the
wavelet modes. The decay is clearly exponential with a decay constant
$\Gamma=0.18$. This corresponds to the theoretical value
$\Gamma=\frac{1}{2}D^2\omega^2$.
\vspace{-0.5cm}

\section{CONCLUSION}

In this paper we have shown how wavelets can be used as basis functions
in canonical quantization. Different mode functions for electric
and magnetic fields, which are localized both in real and Fourier
space, are obtained. Mode functions as well as new operators in
wavelet basis are linear transforms of plane wave mode functions
and operators. The new annihilation and creation
operators satisfy bosonic commutation relations. Because the formalism
remains the same, it is easy to change the basis from plane waves
to wavelets. We have applied the theory to a few example simulations
and showed that the new basis gives well known results in all cases.
In this paper we have used wavelet basis for bosonic
operators in canonical quantization. The same methods can be used to
change basis also for fermionic operators. A localized basis is beneficial
for many solid state and semiconductor physics problems.

There are several generalizations and improvements of the theory described
in this paper. Complex and biorthogonal wavelets \cite{cdf} have some
benefits compared
to real and orthonormal wavelets, for example they can be symmetric.
One generalization is to use wavelet packets or multiwavelets instead
of wavelets. Finally it
would be interesting to compare characteristics of mode functions
of different wavelets. It is also possible to construct new wavelets,
which have desirable properties, for different problems.

\section{ACKNOWLEDGEMENTS}

We thank the Academy of Finland (project 43336) for financial support.
Computers of
the Center for Scientific Computing (CSC) were used in the simulations.
The C++ class library 'blitz' developed by
Todd Veldhuizen was used (http://oonumerics.org/blitz/). We thank
A. R. Baghai-Wadji, N. L\"utkenhaus and K.-A. Suominen for discussions
and comments. More figures of different wavelet mode functions can be found
from the page http://tftsg6.hip.helsinki.fi/\~{ }mhavukai/wavelets/ .

\end{multicols}

\end{document}